\documentclass[preprint]{aastex}
\usepackage{graphics}

\newcommand{\kms}{${\rm km \: s^{-1}}$}

\slugcomment{Submitted to Astrophys. J.}

\shortauthors{Y. Takeda et al.}
\shorttitle{OXYGEN ABUNDANCES IN VERY METAL-POOR GIANTS}

\begin{document}


\title{OXYGEN ABUNDANCE DETERMINATION IN VERY METAL-POOR GIANTS: 
PERMITTED O~I LINES VERSUS FORBIDDEN [O~I] LINES}


\author{Y{\scshape oichi} T{\scshape akeda}}
\affil{Institute of Astronomy, The University of Tokyo,
    Mitaka, Tokyo, Japan 181-0015}
\email{takedayi@cc.nao.ac.jp}

\author{M{\scshape asahide} T{\scshape akada}-H{\scshape idai}}
\affil{Research Institute of Civilization, Tokai University,
1117 Kitakaname, Hiratsuka, Kanagawa, Japan 259-1292}
\email{hidai@apus.rh.u-tokai.ac.jp}

\author{S{\scshape hizuka} S{\scshape ato}}
\affil{Department of Aeronautics, School of Engineering, Tokai University,
1117 Kitakaname, Hiratsuka, Kanagawa, Japan 259-1292}
\email{shizuka@apus.rh.u-tokai.ac.jp}

\author{W{\scshape allace} L. W. S{\scshape argent} and L{\scshape imin} L{\scshape u}}
\affil{Astronomy Department, California Institute of Technology,
Mail Stop 105-24, Pasadena, CA~91125}
\email{wws@astro.caltech.edu}

\author{T{\scshape homas} A. B{\scshape arlow}}
\affil{Infrared Processing and Analysis Center, California Institute of
Technology, Mail Stop 100-22, Pasadena, CA~91125}
\email{tab@ipac.caltech.edu}

\and

\author{J{\scshape un} J{\scshape ugaku}}
\affil{Research Institute of Civilization, 2-29-3 Sakuragaoka, 
Tama-shi, Tokyo, Japan 206-0013}
\email{jugakujn@cc.nao.ac.jp}


\begin{abstract}
The abundance of oxygen was determined for selected very metal-poor
G--K stars (six giants and one turn-off star) based on the high S/N
and high-resolution spectra observed with Keck HIRES in the red 
through near-IR region comprising the permitted O~{\scshape i} lines (7771--5, 8446)
along with the [O~{\scshape i}] forbidden line at 6363 $\rm\AA$. It turned out that both 
the abundances from the permitted line features, O~{\scshape i} 7771--5 and O~{\scshape i} 8446,
agree quite well with each other, while the forbidden line yields
somewhat discrepant and divergent abundances with a tendency of 
being underestimated on the average. 
The former (7773/8446) solution, which we believe to be more reliable, 
gives a fairly tight [O/Fe] vs. [Fe/H] relation such that increasing 
steadily from [O/Fe] $\sim$ 0.6 (at [Fe/H] $\sim$ $-1.5$) 
to [O/Fe] $\sim$ 1.0 (at [Fe/H] $\sim$ $-3.0$), in reasonable consistency
with the trend recently reported based on the analysis of the UV OH lines. 
We would suspect that some kind of weakening mechanism may occasionally 
act on the formation of [O~{\scshape i}] forbidden lines in metal-poor stars.
Therefore, [O~{\scshape i}] lines may not be so a reliable abundance indicator 
as has been generally believed.

\end{abstract}




\section{INTRODUCTION}

The hot controversy on the trend of the oxygen abundance in very old 
metal-deficient stars is one of the most interesting topics in  
the field of stellar spectroscopy as well as Galactic chemical 
evolution. (See, e.g., Carretta, Gratton, \& Sneden 2000 and 
references therein for overviewing the current status of this problem.) 

The essential point of the argument is, shortly speaking, how 
the [O/Fe] ratio behaves itself toward the regime of very low 
metallicity, say, down to [Fe/H]$\sim -3$.
Admittedly, it seems to have been almost established that [O/Fe] increases 
from $\sim 0.0$ ([Fe/H]$\sim 0$) to $\sim 0.4$ ([Fe/H]$\sim -1$)
for stars of disk population.
However, when stars of old halo population are concerned, we are 
wondering whether [O/Fe] shows a nearly flat plateau at [O/Fe] $\sim +0.5$ 
(``plateau'' trend) or  it continues to increase steadily up to 
[O/Fe] $\sim +1$ (``steady increase'' trend), as [Fe/H] decreases 
from $\sim -1$ down to $\sim -3$.

From observational point of view, this problem is closely connected 
with the technical details of abundance determination; i,e., especially
important is which of the spectral lines to be adopted.
In the present case of late-type stars, the candidates of the lines used for 
O-abundance determination are roughly divided into three groups : \\
(1) O~{\scshape i} permitted lines (e.g., 7771--5 triplet)\\
(2) [O~{\scshape i}] forbidden lines (6300/6363 doublet)\\
(3) OH lines in the UV region of 3100--3200 $\rm\AA$ \\
Among these the [O~{\scshape i}] forbidden lines (originating from the ground level) 
have so far been regarded as being the most reliable abundance indicator
because of its low $T_{\rm eff}$ sensitivity along with the certain validity 
of the LTE assumption, in contrast to the high-excitation permitted lines
which are $T_{\rm eff}$-sensitive and more or less affected by a non-LTE effect.

Generally speaking, there seems to have been an implicit consensus that
the ``plateau'' trend probably represents the truth, partly because 
it is suggested from the analysis of [O~{\scshape i}] 6300 forbidden line,
though with a rather large scatter (see, e.g., Fig. 11 in 
Timmes, Woosley, \& Weaver 1995). 
In addition,  the fact that this ``flat'' behavior is consistent with the trend
posed by other $\alpha$-elements (Mg, Si, Ca, Ti) may have substantiated
this view. Namely, it is reasonably explainable based on the 
standard theory of Galactic chemical evolution. That is, a steady enrichment 
of $\alpha$-elements along with Fe with a constant $\alpha$-to-Fe ratio 
by massive stars via type II SNe 
(very short time-scale nearly like instantaneous recycling)
in the earlier phase of Galactic evolution, and then later 
type Ia SNe (long time-scale of the order of $10^{9}$ yr) begins to 
explode to eject Fe, which must have gradually suppressed [$\alpha$/Fe] 
eventually down to the current value.
 
On the other side, the O~{\scshape i} 7771--5 lines were known to imply 
the ``steady increase'' trend (see, e.g., Fig. 1 in Bessell, Sutherland, 
\& Ruan 1991, or Fig. 4 in Mishenina et al. 2000), 
which means that these permitted lines
yield higher abundance than that resulted from the forbidden line.
Nevertheless, it appears that this tendency has not been taken 
too seriously compared to the [O~{\scshape i}] results for the uneasy problems 
involved with these lines mentioned above.

Recently, however, Israelian, Garc\'{\i}a L\'{o}pez, \& Rebolo (1998) 
and  Boesgaard et al. (1999)
reported that [O/Fe] clearly increases from $\sim$+0.6 to $\sim$+1.0 as [Fe/H] 
varies from $\sim$ $-1.5$ to $-3$ (just similar to what had been implied
by 7771--5 permitted lines) based on their new independent analyses 
of numerous OH lines in UV, which have sufficient strengths even in 
extremely metal-poor stars and thus suitable for abundance determination.
This ``steady increase'' trend has thus acquired a special attention
as the true [O/Fe] vs. [Fe/H] relation,
though this must be quite embarrassing for theoreticians
(i.e., theoretical interpretation becomes much harder).

Even so, the problem is still far from being settled, unfortunately. 
A strong criticism was successively addressed by Fulbright \& Kraft (1999),
who reinvestigated the atmospheric parameters of two extremely
metal-poor stars sudied by Israelian et al. (1998) and showed
that their large [O/Fe] values ($\sim +1$) can be considerably 
reduced down to $\sim +0.5$ if the [O~{\scshape i}] 6300 line was used,
concluding that it is still premature to accept their OH results.

So we are in a quite confusing situation. Which line gives the correct
oxygen abundance?  [O~{\scshape i}] forbidden line?  
O~{\scshape i} permitted lines 
(which yield the same trend as that given by OH lines)? 
This topic attracted our attention, since we once confirmed that 
the oxygen abundances derived from O~{\scshape i} 7771--5 lines and 
[O~{\scshape i}] 6300 line 
do not show any systematic discordance from the analysis of Population I 
G--K giants (Takeda, Kawanomoto, \& Sadakane 1998a).

Consequently, we decided to conduct a detailed analysis of selected seven
very metal-deficient stars ($-3$ $\la$  [Fe/H] $\la$ $-1.5$)
based on the high-quality spectra observed with Keck HIRES, 
in an attempt to compare the abundances of oxygen derived from
the permitted O~{\scshape i} lines (7771--5, 8446) and forbidden [O~{\scshape i}] lines (6363, 6300) 
with each other, as well as to settle which of the [O/Fe] trend 
is more likely. We will show later that the former ``permitted'' O~{\scshape i} lines
are probably a more reliable abundance indicator rather than the forbidden
[O~{\scshape i}] lines, lending support for the ``steadily increasing'' trend of [O/Fe]. 

\section{OBSERVATIONAL DATA}

\subsection{Observations}

The spectroscopic observations were carried out for six HD-numbered
very metal-deficient stars (five G--K giants and one turn-off star in 
the metallicity range of $-3$ $\la$ [Fe/H] $\la$ $-1.5$)
in the seasons of 1997 May and 1999 November using the 
High Resolution Echelle Spectrograph (HIRES; Vogt 1994) of
the Keck I Telescope. Basic observational data of the program
stars are given in Table 1.

\subsection{Data Reduction}

The data were reduced with the HIRES  Automated Reduction (HAR)
software package developed by T. A. Barlow. This package converts the 
two-dimensional echelle raw data to one-dimensional wavelength-calibrated
spectrum of each order with full processings of bias subtraction,
flat fielding, cosmic-ray correction, sky subtraction, and dispersion 
correction. 
The multiple calibrated spectra  were coadded using the IRAF package
{\it scombine} and then normalized to 
continuum level of unity using the IRAF package {\it continuum} for  1997 May 
data and the HAR package {\it xplot} for 1999 November data.   

The wavelength range of the spectrum
is 6330 $\rm\AA$ $\la$ $\lambda$ $\la$ 8760 $\rm\AA$, 
which comprises the lines of [O~{\scshape i}] 6363, O~{\scshape i} 7771--5, and O~{\scshape i} 8446.
Since we tried to cover the long-wavelength range up to 
8760 $\rm\AA$ (because of our intention of observing special spectral 
lines such as those of sulfur), we were obliged to set the short-wavelength 
limit at 6330 $\rm\AA$. Consequently, we had to abandon using the [O~{\scshape i}] 6300 
line.  

\subsection{CS 22892-052}

Regarding CS 22892-052, the well-known carbon-rich very metal-poor star 
with excesses of neutron-capture elements
(cf. Sneden et al. 1994), we used the Keck HIRES spectrum 
which was kindly reduced and provided by M. Kuchner (private communication)
covering the wavelength range of 
5340$\rm\AA$ $\la$ $\lambda$ $\la$ 7770 $\rm\AA$.
Unfortunately, the signal-to-noise ratio of this spectrum is considerably 
poor (120--140) compared to those of other six stars.
Since the long-wavelength limit is just on the position of O~{\scshape i} 7771--5 
triplet, only the shortest-wavelength component (at 7771.94 $\rm\AA$) 
of the three could be used. Note also that the [O~{\scshape i}] 6300 line was usable 
in addition to [O~{\scshape i}] 6363, while the O~{\scshape i} 8446 line is not included in 
this spectrum. 

\subsection{Equivalent-Width Measurements}

The identification of spectral lines, the selection of blend-free lines
usable for abundance determination, and the evaluation of equivalent widths
by Gassian fitting (or by direct integration if necessary) 
were carried out with the help of the interactive GUI software ``SPSHOW'',
which was developed by Y. Takeda based on Kurucz's ATLAS/WIDTH code 
as a multi-purpose tool for stellar spectrum analysis (e.g., synthesizing 
a theoretical spectrum with any abundance/line-broadening,
browsing of observed spectrum to be overplotted, radial-velocity correction
by fitting with an appropriate horizontal shift, interactive 
abundance-determination, equivalent-width measurement, etc.)
We measured the equivalent width for all spectral lines as possible
(irrespective of the species) as far as being confidently identified 
and free from blending with other lines. These data were further used for
abundance determination as described in \S 3.2. 

\section{ABUNDANCE DETERMINATION}

\subsection{Model Atmospheres}

We used Kurucz's (1993) ATLAS9 line-blanketed model atmospheres,
based on which the atmospheric models of individual stars were constructed
by interpolation in terms of $T_{\rm eff}$, $\log g$, and [X] (metallicity).

Regarding the choice of these parameters, we adopted the values established 
by Pilachowski, Sneden, \& Kraft (1996) if available (HD 44007, HD 165195, 
HD 175305, HD 184266), since their determinations appear to be reliable,
because they were based on careful considerations and checked from 
various standpoints.

As for HD 84937 and HD 88609, the $T_{\rm eff}$ values were evaluated
from $(b-y)_{0}$ color (0.30 and 0.65), which were taken from 
the compilation of Hauck and Mermilliod (1998) and corrected for reddening
as described below, in the same way as was done by Pilachowski et al. (1996), 
yielding 6040K and 4440K, respectively. 
We applied a slight reddening correction ($-0.03$) corresponding to the 
${\rm E}_{B-V}$ value of +0.04 (cf. Bond 1980; ${\rm E}_{b-y}$ = 
0.73${\rm E}_{B-V}$) for HD 88609 to obtain $(b-y)_{0}$ from the raw $b-y$,
while ${\rm E}_{b-y}$ = 0 could be 
safely assumed for the nearby (81 pc) turn-off star HD 84937.
We then computed the $\log g$ values from $L$ (bolometric 
luminosity), which were derived from Hipparcos data with the 
bolometric correction estimated from Kurucz's (1993) theoretical colors,
while using the adopted $T_{\rm eff}$ and assuming 0.8 $M_{\sun}$ 
(Pilachowski et al. 1996), and eventually obtained 4.0 and 0.8, respectively.  
Comparing such calculated  $T_{\rm eff}$ and 
$\log g$ with the published values compiled by Cayrel de Strobel et al. (1997),
which we also consulted for estimating the stellar metallicity,
we finally adopted (6100K, 4.0, $-2.3$) and (4500K, 1.0, $-2.7$)
for ($T_{\rm eff}$, $\log g$, [X]) of HD 84937 and HD 88609, respectively.

Concerning CS 22892-052, we simply used the parameters determined by 
McWilliam et al. (1995). The finally adopted model parameters for
our program stars are given in Table 2.

\subsection{Microturbulence and Elemental Abundances}

We used the WIDTH9 program written by R. L. Kurucz 
to determine the elemental abundances,
where the $gf$ values compiled by Kurucz (1995) were exclusively used.

Practically, the values of the microturbulence were first established 
from the equivalent widths of Fe~{\scshape i} lines, following the procedure 
adopted by Takeda (1992), and are given in Table 2.
We then determined the abundances of various elements
(apart from C and O which were treated specially)
from the measured equivalent-widths of spectral lines,
with the corresponding atmospheric model and the microturbulent velocity.

The resulting abundances averaged over each of the lines are presented 
in Table 2. In case where lines of two ionization stages were available, 
we took their simple mean.

Although the detailed data and results for each line (e.g., the equivalent 
width, the $gf$ value, and the abundance, etc. ) are not shown in this paper, 
complete tables in a machine-readable form containing all these data are 
downloadable from the following anonymous ftp site.
\begin{verbatim}
IP address: 133.11.160.7 
directory:  /Users/takeda/oxygen2000/tables/
\end{verbatim}
We confirmed almost satisfactory accomplishment of the ionization 
equilibrium between Fe~{\scshape i} and Fe~{\scshape ii}, since the abundance difference 
(Fe~{\scshape i} $-$ Fe~{\scshape ii}) turned out to be  +0.08, $-0.09$, $-0.14$, $-0.07$, +0.17, 
+0.05, and +0.06 (in the same order as in Table 2).
Note also that the resulting [Fe/H] values are consistent with the
adopted metallicity of the model.

\subsection{Oxygen Line Analysis}

\subsubsection{Non-LTE Calculation}

The statistical-equilibrium calculations for oxygen were carried out 
twice for each star by assuming two input oxygen abundances [O/Fe] = 0 and 1.
See Takeda et al. (1998a) and the references therein for computational 
details.
The resulting departure coefficients and line-source functions were further
used to determine the non-LTE oxygen abundance.

\subsubsection{CO Molecule Effect}

As described in the following, we determine the oxygen abundance
from the equivalent widths of oxygen lines or by profile-fitting.
Here, there is one thing we should be aware in the present case of 
late-type stars.
Since an appreciable fraction of C and O atoms are combined to form
CO molecules, the determination of the abundance of oxygen more or less 
depends upon that of C along with the O-abundance itself 
(i.e., the abundance adopted in the model atmosphere).
In order to take this effect properly into consideration,
we determined the abundance of C for our program stars
from [C~{\scshape i}] 8727.13 (HD 44007, 88609, 165195) or 
C~{\scshape i} 8335.15 (HD 84937, 175305, 184266), except for CS 22892-052.

Practically, we first determined the O-abundance by assuming
[C/Fe] = 0 ([C/Fe] = +1 only for CS 22892-052; cf. McWilliam et al. 1995).
With such determined abundance of oxygen, the abundance of C was
evaluated from the equivalent width of the [C~{\scshape i}] or C~{\scshape i} line.
Finally, the oxygen abundance was again computed with this updated
C-abundance.

Actually, this whole procedure was repeated twice in order to derive two
oxygen abundances $\log\epsilon_{0}$(O) and $\log\epsilon_{1}$(O),
corresponding to two different input oxygen abundances of the model 
atmosphere, [O/Fe] = 0 and +1. Then, the final oxygen abundance to be adopted
was derived by finding the consistent solution based on these two.

Nevertheless, both C-dependence and the (input) [O/Fe]-dependence were
found to be generally insignificant and negligible in most cases 
(a few hundredths dex at most). The resulting carbon abundances are
given in Table 2.

\subsubsection{O~I 7771--5}

The abundances from these well-known high-excitation O~{\scshape i} triplet lines 
(RMT No.1; $\chi$ = 9.15 eV) in the quintet system 
($3s\; ^{5}S^{o}$ -- $3p\; ^{5}P$) 
were derived from their equivalent widths. The $gf$ values (and the 
damping parameters) were taken from Kurucz (1995); i.e., +0.32, +0.17, 
and $-0.05$ for the lines at 7771.94, 7774.17, and 7775.39 $\rm\AA$, 
respectively.
The results are given in Table 3, where the mean of the non-LTE abundances
over the components, $<\log\epsilon_{7773}>$, is also presented.
(We will regard this $<\log\epsilon_{7773}>$ as being the standard oxygen 
abundance to be adopted.) The theoretical spectrum computed with this
$<\log\epsilon_{7773}>$ are compared with the observed one for each star
in Figure 1, where we can confirm that the fit is quite satisfactory.

In the case of HD 184266, we found that consistency
of the abundances for these three components could not be attained with the
use of $\xi$ = 2.2 \kms , which was derived from Fe~{\scshape i} lines (\S 3.2). 
We thus adopted the value of 5.1 \kms (the best-fit value established by 
profile-fitting; cf. \S 6.1 of Takeda \& Sadakane 1997) for this star. 
This problem will be discussed further in \S 4.1.

\subsubsection{O~I 8446}

In contrast to the famous O~{\scshape i} 7771--5 triplet, this similar high-excitation 
O~{\scshape i} 8446 feature (RMT No.4; $\chi$ = 9.52 eV) in the triplet system 
($3s\; ^{3}S^{o}$ -- $3p\; ^{3}P$) has barely been used for oxygen abundance
determination, presumably because it is seriously blended with Fe~{\scshape i} lines,
We thus applied a profile-fitting technique (Takeda 1995a) to the analysis
of this O~{\scshape i} + Fe~{\scshape i} 8446 blended feature, such as was done by Takeda et al. 
(1998b). The adopted $gf$ values of the O~{\scshape i} components 
(8446.25, 8446.36, and 8446.76 $\rm\AA$) are ($-0.52$, +0.17, and $-0.05$)
taken from Kurucz (1995), while those of the blended Fe~{\scshape i} lines 
(8446.39 $\rm\AA$, $\chi$ = 4.99 eV and 8446.57 $\rm\AA$, $\chi$ = 4.91 eV)
were the empirical values of ($-0.85$ and $-1.89$) determined by Takeda 
(1995b). The Fe abundance was simultaneously determined along with that of O
as far as possible; otherwise, an appropriate value of the Fe abundance 
corresponding to the metallicity
was assigned and fixed. The resulting oxygen abundances are given in Table 3, 
where the O~{\scshape i} equivalent widths [ $W_{\lambda}$(O~{\scshape i} 8446) ] corresponding to 
the pure contribution of three O~{\scshape i} component lines, which were inversely 
evaluated by spectrum synthesis while using the established solutions of 
the oxygen abundance, are also presented.
With the equivalent widths of two Fe~{\scshape i} lines similarly computed,
we found that the estimated fractions of
$W_{\lambda}$(O~{\scshape i} 8446) / [ $W_{\lambda}$ (O~{\scshape i} 8446) +  
$W_{\lambda}$ (Fe~{\scshape i} 8446.36) + $W_{\lambda}$ (Fe~{\scshape i} 8446.57) ]
turned out to be 81\%, 98\%, 53\%, 72\%, and 98\%, for 
HD 44007, HD 84937, HD 165195, HD 175305, and HD 184266, respectively.
The theoretically synthesized spectra computed with $<\log\epsilon_{7773}>$
are shown in Figure 2. 

\subsubsection{[O~I] 6363/6300}

We adopted Kurucz's (1995) $gf$ value for the [O~{\scshape i}] forbidden lines 
($2p^{4}\; ^{3}P$ -- $2p^{4}\; ^{1}D$); i.e., 
$\log gf$ = $-9.82$ for [O~{\scshape i}] 6300.30 ($\chi$ = 0.00 eV) 
and $-10.30$ for [O~{\scshape i}] 6363.78 ($\chi$ = 0.02 eV),
though the former was usable only for CS 22892-052.
The abundance results are given in  Table 3, while the theoretical
spectra corresponding to $<\log\epsilon_{7773}>$ are shown in Figure 3.
Note that the abundances for CS 22892-052 (6363 and 6300 lines) and 
HD 184266 (6363 line) are uncertain, because of the difficulty
in measuring their equivalent widths accurately (i.e., line profiles 
of these lines are not well defined).

\section{DISCUSSION}

\subsection{HD 184266: Importance of Turbulent Velocity}

Before discussing the trend of the oxygen abundance, it may be worth
paying attention to HD 184266, which shows markedly stronger
O~{\scshape i} lines (7771--5, 8446) than any other star, in contrast to
the remarkable weakness of [O~{\scshape i}] 6363 exhibiting a rather
ambiguous profile. 

As mentioned in \S 3.3.3, the abundances derived from three components of 
the O~{\scshape i} 7771--5 turned out to be discordant with each other
when the microturbulent velocity of 2.2 \kms (based on Fe~{\scshape i} lines)
was used (cf. Table 3). The consistency can be accomplished only when
this turbulence parameter is considerably raised up to $\sim$ 5 \kms .
We point out that this is very similar to the situation experienced by 
Takeda \& Sadakane (1997) in their analysis of Population II
horizontal-branch star HD 161817 (cf. \S 6.1 therein), 
in which they attributed this discrepancy to the depth-dependence
of the turbulent velocity field increasing with atmospheric height

In fact, similar tomography on $\xi$ as was done by Takeda \& Sadakane (1997) 
revealed that two groups of Fe~{\scshape i} lines of
different $\chi$ (lower excitation potential) lead to
discrepant values for the $\xi$ value of HD 184266; i.e., 
2.3 \kms\ with $<\log\tau> \sim -0.6$ (0 eV $\leq \chi <$ 4 eV) and
1.8 \kms\ with $<\log\tau> \sim -0.4$ (4 eV $\leq \chi $),
indicating a tendency of upward-increasing turbulence.
That such a similar phenomenon takes place in the atmosphere of 
both stars becomes even convincing when we recall HD 184266 is
considered to be a horizontal-branch star (i.e., just same as HD 161817)
as judged from its position in the HR diagram 
(cf. Fig. 3 in Pilachowski et al. 1996). 
Hence we may state that the atmospheric turbulent velocity field
in these horizontal-branch stars grows with an increase in
height, significantly affecting the strength of
such high-excitation O~{\scshape i} lines, because they become fairly strong 
due to relatively high $T_{\rm eff}$ and tend to form higher in the 
atmosphere. 

It is not clear whether such a phenomenon has something to do
with the metal-deficiency. Yet, the recent work by
Gadun \& Ploner (1999) is interesting in this connection, since
their calculation indicates that a decrease in metallicity causes
the following effects in stellar atmospheres.\\
-- A growth of granulation contrast.\\
-- A growth of atmospheric oscillation.\\
-- An increase of temperature variation in the upper photosphere.\\
-- Size of granules tends to shift toward smaller scales.\\
Accordingly, we had better pay attention to the treatment of 
turbulent velocity fields in the atmosphere of metal-poor stars
more seriously, at least when saturated lines are concerned.

\subsection{Permitted vs. Forbidden Lines}

We now discuss the main topic of this paper; i.e., comparison of the
oxygen abundances derived from different lines. 

As can be seen from Figure 4, the abundances derived from both permitted 
line features of high excitation (O~{\scshape i} 7771--5 and O~{\scshape i} 8446) are in 
remarkable agreement with each other, which can be also intuitively 
confirmed from Figure 2. This means that these two abundances may be regarded
as being essentially equivalent, and that O~{\scshape i} (+ Fe~{\scshape i}) 8446 is practically
usable for oxygen abundance determination supplementing O~{\scshape i} 7771--5.

On the other hand, the case of the forbidden line ([O~{\scshape i}] 6363 or 6300)
is more complicated. What we notice from Figure 4 is the large scatter
in the 6363(6300) vs 7773 correlation. More precisely, while a good agreement
is observed for HD 165195 and HD 175305, the forbidden [O~{\scshape i}] line tends
to yield lower abundance (typically by 0.3--0.4 dex) than those determined 
from the permitted O~{\scshape i} lines for other stars.

According to Table 3, if the $T_{\rm eff}$ of the model atmosphere were 
increased (say, by $\sim 200$ K), it would be possible to bring them into 
consistency, while the the effect of changing $\log g$ acts in the same 
direction for both O~{\scshape i} and [O~{\scshape i}] lines. Yet, we would consider
it rather unlikely that the $T_{\rm eff}$ values of our model atmospheres are 
correct for some stars on the one hand but not for others on the other, 
since our adopted $T_{\rm eff}$ scale 
is based on a consistent system of Pilachowski et al. (1996) 
(cf. \S 3.1).

Hence, we regard this discrepancy as being real, in the sense that 
it is unlikely to be attributed to any systematic error in abundance 
determination. 
Accordingly, in contrast to Population I G--K giants
where the abundances from O~{\scshape i} 7773 lines and [O~{\scshape i}] 6300 do not show
any systematic discordance (Takeda et al. 1998a), it may be stated that 
``{\it metal-poor Population II} G--K giants tend to exhibit abundance 
discrepancy between these two O~{\scshape i} and [O~{\scshape i}] lines, though this does not 
necessarily always happen (i.e., there are surely cases where the consistency 
is achieved).''  

\subsection{[O/Fe] vs. [Fe/H] Relation}

The resulting [O/Fe] vs. [Fe/H] correlation for O~{\scshape i} 7771--5 and 
[O~{\scshape i}] 6363(6300)  is depicted in Figure 5, where Israelian et al.'s (1998)
results based on the OH lines are also shown for comparison. 
It can be seen from this figure that the [O/Fe] values from O~{\scshape i} 7771--5 lines 
are reasonably consistent with the OH results showing a ``steady increase''
trend, while those from [O~{\scshape i}] forbidden lines tend to lie below 
the O~{\scshape i} or OH plots by several tenths dex {\it with a considerably
large scatter}. In any case, we have reconfirmed the observational 
disagreement between the [O/Fe] vs. [Fe/H] relation derived 
from O~{\scshape i} and [O~{\scshape i}] lines mentioned in \S 1.

Then, which represents the truth?
We have a preference for the O~{\scshape i} permitted lines (7771--5/8446),
because they show a clean trend consistent with that exhibited by OH lines.
The large diversity of the [O/Fe] values indicated by [O~{\scshape i}] lines
gives us a feeling that these forbidden lines may be unreliable. 
Consequently, we conclude that the [O/Fe] vs. [Fe/H] relation for 
metal-poor stars has the ``steady increase'' trend, based on our 
permitted O~{\scshape i} lines (7711--5/8446) analysis, lending support 
for the OH results reported by Israelian et al. (1998) and 
Boesgaard et al. (1999). 

Turning our attention to other recent works, we note that
Mishenina et al. (2000) also arrived at the same conclusion as ours based on 
their non-LTE analysis of the permitted O~{\scshape i} 7771--5 triplet,
as suggested by previous works (cf. \S 1).
However, there are reports that even the permitted O~{\scshape i} lines 
yield the ``flat''-like trend similarly to [O~{\scshape i}] lines 
(e.g., Carretta et al. 2000; Fig. 1 in Gustafsson 1999).
The reason for this discordance is not clear;
further observational work on a much larger sample of very metal-poor stars
and analysis in a fairly consistent system will be needed until this
confusion is finally settled.

\subsection{Weakening of [O~I] line?}

Our choice (for the preference of O~{\scshape i} and OH lines) naturally means 
that the abundances from [O~{\scshape i}] lines tend to be erroneously underestimated
for some reasons. What has happened to these forbidden lines ?

We suspect that this line may often suffer with an appreciable weakening
in a non-classical way. As a matter of fact, the possibility of such a 
weakening was once remarked by Langer (1991), who proposed a model 
that the absorption trough of the low-excitation [O~{\scshape i}] forbidden line 
may be filled with the [O~{\scshape i}] emission from the surrounding nebulosity,
thus leading to a weakening of the line. 
In this connection, it is 
interesting to refer to the [O~{\scshape i}] 6363 line of HD 184266,
which shows a large O~{\scshape i} vs. [O~{\scshape i}] discordance amounting to $\sim$ 0.5 dex 
(cf. Table 3). Let us pay attention to Figure 3, where we notice that
the line profile of the [O~{\scshape i}] 6363 line is rather complex and a kind of
emission-like feature appears to exist (though not definite). 
Could it be an emission from the surrounding nebulosity, which, for example,
was formed by mass ejection at the RGB phase in the past (recall that 
this is a horizontal-branch star)? 
Unfortunately, this line is too weak to make a definite argument 
even with this high S/N ratio, and it is also probable that what we see
is nothing but noises. 
At any rate, we consider that Langer's (1991) mechanism may be worth
further investigation. Such a dilution mechanism due to emission lines
(if any exists)
should be relatively more important for cases where the strength of the 
[O~{\scshape i}] absorption line is intrinsically weak (i.e., more efficient for
metal-poor case), which might explain the reason why the discordance is 
observed only in Population II stars (but not in Population I).

In addition, we should not forget the possibility that the effect of 
atmospheric inhomogeneity may cause appreciable descrepancies of abundances
derived from different lines when analyzed classically, which can be tackled  
only by detailed realistic (2D or 3D) modeling of the atmosphere of 
metal-poor stars (e.g., Asplund et al. 1999).

\subsection{Comparison with Other $\alpha$-Elements}

We thus conclude that [O/Fe] steadily increases from $\sim$+0.6 to
$\sim$+1.0 as the metallicity becomes lower from [Fe/H] $\sim -1.5$
down to [Fe/H] $\sim -3$. 

Frankly speaking, however, this must be a rather embarrasing trend
hard to interpret for theoreticians who are constructing models of 
Galactic nucleosynthesis history, because other $\alpha$-elements 
(Mg, Si, Ca, Ti), which are believed to form in massive stars and ejected by
type II SNe explosion like oxygen, generally show more or less
the trend of flat ``plateau'' (i.e., [$\alpha$/Fe] $\sim$ 0.5 at
[Fe/H] $\sim -3$; cf. Fig. 2 in Ryan, Norris, \& Beers 1996).
This can be confirmed also from the results of Mg, Si, Ca, and Ti 
derived for our program stars (cf. Table 2).
This would imply that some kind of unusual enrichment (e.g., pre-Galactic
origin, or by supermassive stars) may have taken place for oxygen 
(while not for Mg, Si, Ca, and Ti) at the very early phase of the Galaxy, 
which would have caused an exceptionally large [O/Fe] ratio at the old time 
of very low metallicity, though we are not qualified to state about 
wherether such a complicated modeling is possible or not.

Finally, we make one interesting remark concerning sulfur,
another $\alpha$-element which has not been well studied so far.
According to our results given in Table 2, 
which were determined from only one S~{\scshape i} line at 8694.63 $\rm\AA$, 
the [S/Fe] ratio shows a steeply increasing trend 
(from [S/Fe] $\sim$ +0.3 at [Fe/H] $\sim -1.5$
to [S/Fe] $\sim$ +1.1 at [Fe/H] $\sim -2.7$), which is similar to the
tendency of [O/Fe] but not to that of [Mg/Fe], [Si/Fe],
[Ca/Fe], and [Ti/Fe]. Although this kind of behavior for [S/Fe]
(as well as for [O/Fe]) is not consistent with the current
theoretical prediction (e.g., Timmes et al. 1995), their similarity may 
be worth attention, which might be a key to understanding the difference 
between these two groups of $\alpha$-elements. 
More detailed analysis and discussion on S will be published 
in a separate paper (Takada-Hidai et al., in preparation).

\section{CONCLUSIONS}

We determined the oxygen abundance (along with the abundances of other 
elements) by a detailed non-LTE analysis for selected seven very 
metal-poor stars of 1/30 $\sim$ 1/1000 solar metallicity, based on 
the high-S/N  spectra (in the red through near-IR region) 
observed with Keck HIRES, in order to 
study whether any significant systematic difference exist between the 
abundances derived from high-excitation O~{\scshape i} permitted lines (7771--5, 8446) 
and those from low-excitation [O~{\scshape i}] forbidden lines (6363, 6300).

The relatively hot ($T_{\rm eff}$ = 5600 K) horizontal-branch star
HD 184266 turned out to be an interesting object, 
which shows markedly stronger O~{\scshape i} permitted lines 
(7771--5, 8446) and very weak [O~{\scshape i}]  forbidden lines (6363, 6300).
It revealed that the microturbulent velocity dispersion increases
with height in the atmosphere of this star: this implies that 
one should take care about the choice of microturbulence
in determining the abundance from saturated lines such as those
permitted O~{\scshape i} lines.
Since the rather complex profile of [O~{\scshape i}] 6363 line might be a sign of
emission component (though not definite), Langer's (1991) hypothesis
(diluton due to the ``filled-in'' emission from the circumstellar nebulosity)
may deserve special attention as a possible weakening mechanism
for this forbidden line.

We found that the both O~{\scshape i} 7771--5 and O~{\scshape i} 8446 yield quite consistent
abundances with each other. On the other hand, the abundance
determined from [O~{\scshape i}] 6363 (or 6300) is generally lower typically by a few 
tenths dex than that derived from O~{\scshape i} permitted lines,
though the extent of the difference varies from star to star 
(two stars show a reasonable consistency).

The resulting [O/Fe] vs. [Fe/H] relation based on the permitted O~{\scshape i} 7771--5
triplet shows a tight and clear tendency that [O/Fe] increases steadily 
from $\sim$+0.6 to $\sim$+1.0 as [Fe/H] decreases from $\sim -1.5$
down to $\sim -3$, which is quite similar to the trend found by the recent 
analyses of OH lines in UV (Israelian et al. 1998; Boesgaard et al. 1999).
On the other hand, the [O/Fe] values derived from [O~{\scshape i}] lines,
which tend to be lower than the results of O~{\scshape i} permitted lines 
by $\sim$ 0.3 dex on the average, show a large diversity and considered 
to be unreliable compared to what was suggested from permitted O~{\scshape i} lines.

Consequently, we would conclude that the most probable [O/Fe] vs. [Fe/H]
relation at the very metal-deficient regime is the ``steady increase''
trend implied by O~{\scshape i} permitted lines and OH lines in UV.
We suspect that the [O~{\scshape i}] forbidden lines tend to suffer with
some kind of weakening, yielding an underestimated abundance,
which becomes appreciable only in Population II metal-poor stars.
Thus, according to our opinion, its use for abundance determination is
questionable, in contrast to the general belief.
All what we can state about the possible weakening mechanism is 
only speculative. Its cause might exist outer than the photosphere 
(e.g., nebulosity emission); alternatively, the photosphere itself 
might be responsible for the weakening (e.g., atmospheric inhomogeneity).

The [O/Fe] vs. [Fe/H] trend concluded in this paper can not be 
explained by the standard scenario of Galactic chemical evolution,
since this does not conform to the tendency suggested from other 
well-studied $\alpha$-elements such as Mg, Si, Ca, and Ti.
According to the by-product of this study, however, sulfur 
(another $\alpha$ element) revealed an interesting behavior, in
the sense that [S/Fe] shows a similar tendency to [O/Fe].
We hope that this observational implication will inspire the 
motivation of theoreticians toward a construction of better and 
realitic modeling of Galactic chemical evolution.



\acknowledgments

We are grateful to M. Kuchner for the reduction of the Keck HIRES
data of CS 22892-052 as well as for putting the reduced
spectrum at the authors' disposal.

One of us (MTH) acknowledges the financial supports from grant-in-aid
for the scientific research (A-2, No. 10044103) by Japan Society
for the Promotion of Science as well as from grant-in-aid for research
and education by Tokai University in the 1999 fiscal year, 
which enabled his observation with HIRES in 1999.

This research has made use of the SIMBAD database, operated at CDS, 
Strasbourg, France.





\clearpage
\setcounter{table}{0}
\begin{table}[h]
\caption{O{\scshape bservational} D{\scshape ata}}
\begin{center}
\begin{tabular}{lrllcccc}\hline\hline
 Star  &   ~~$V$~~  &  Sp.Type &  Date (UT)&  No. of &    Total &     R   &    S/N \\
       &  &  &  &    Exposure  & Exposure(s) &  & \\
\hline
HD 44007  & 8.06 & G5IV:w & 1999 Nov 11   &  3  &     900 &     60000 &   200-310 \\
HD 84937  & 8.28 & sdF5   & 1999 Nov  9   &  3  &     900 &     60000 &   190-360 \\
HD 88609  & 8.64 & G5IIIw & 1999 Nov 10   &  3  &     900 &     60000 &   230-340 \\
HD 165195 & 7.34 & K3p    & 1997 May 29-30 & 10 &      820 &     45000 &    370-510 \\
HD 175305 & 7.20 & G5III  & 1997 May 10   &  2  &     400  &    45000  &  410-660 \\
HD 184266 & 7.57 & F2V    & 1997 May 28   &  3  &     600  &    45000  &  330-550 \\
\hline
\end{tabular}
\end{center}
N{\scshape ote} --- Apparent visual 
magnitudes ($V$) and spectral types were taken from the SIMBAD database.
\end{table}

\clearpage
\begin{table}[h]
\caption{A{\scshape dopted} M{\scshape odel} P{\scshape arameters} 
{\scshape and} T{\scshape he} R{\scshape esulting} A{\scshape bundances}}
\begin{center}
\scriptsize
\begin{tabular}{crrrrrrrr}\hline\hline
Star & 44007 & 84937 & 88609 & 165195 & 175305 & 184266 & 22892-52 & ~~~Sun \\
\hline
 &  &  &  &  &  &  &  &  \\
$T_{\rm eff}$ (K) & 4850 & 6100 & 4500 & 4450 & 5100 & 5600 & 4760 &  \\
$\log g$ (${\rm cm}\, {\rm s}^{-2}$) & 2.00  & 4.00  & 1.00  & 1.10  & 2.50  & 1.70  & 1.30  &  \\
$[$X$]$ (metallicity) & $-$1.50  & $-$2.30  & $-$2.70  & $-$2.00  & $-$1.50  & $-$1.70  & $-$3.10  &  \\
$\xi$ (${\rm km}\, {\rm s}^{-1}$) & 1.4 & 1.1 & 1.9 & 1.8 & 1.5 & 2.2 & 1.9 &  \\
 &  &  &  &  &  &  &  &  \\
\hline
 &  &  &  &  &  &  &  &  \\
C & 6.98  & 6.71  & 6.44  & 6.02:  & 7.17  & 6.94  & $\cdots$ & 8.56  \\
O & 7.96  & 7.51  & 7.01  & 7.52  & 7.80  & 7.92  & 7.02  & 8.93  \\
Na & $\cdots$ & $\cdots$ & $\cdots$ & $\cdots$ & $\cdots$ & $\cdots$ & 3.78  & 6.33  \\
Mg & 6.39  & 5.64  & 5.43  & 5.69  & 6.45  & 6.32  & 5.13  & 7.58  \\
Al & 4.67  & $\cdots$ & $\cdots$ & $\cdots$ & $\cdots$ & $\cdots$ & $\cdots$ & 6.47  \\
Si & 6.23  & 5.96  & 5.22  & 5.76  & 6.39  & 6.39  & $\cdots$ & 7.55  \\
S & 5.92 & 5.87 & 5.60: & 5.68 & 6.07 & 5.98 & $\cdots$  & 7.21  \\
K & $\cdots$ & $\cdots$ & $\cdots$ & $\cdots$ & $\cdots$ & $\cdots$ & 2.58 & 5.12  \\
Ca & 5.08  & 4.47  & 3.73  & 4.28  & 5.20  & 5.07  & 3.71  & 6.36  \\
Sc & 1.72  & $\cdots$ & 0.22  & 1.21  & 1.94  & 1.61  & $-$0.10  & 3.10  \\
Ti & 3.51  & 3.37  & 2.37  & 2.94  & 3.83  & 3.47  & 2.25  & 4.99  \\
Cr & 3.97  & $\cdots$ & 2.90  & 3.32  & 4.25  & 4.33  & 2.44  & 5.67  \\
Mn & 3.37  & $\cdots$ & $\cdots$ & $\cdots$ & 4.06  & $\cdots$ & $\cdots$ & 5.39  \\
Fe & 5.94  & 5.34  & 4.79  & 5.46  & 6.18  & 5.94  & 4.61  & 7.51  \\
Co & 3.33  & $\cdots$ & $\cdots$ & 2.69  & 3.67  & $\cdots$ & $\cdots$ & 4.92  \\
Ni & 4.65  & 3.93  & 3.42  & 4.05  & 4.89  & 4.62  & 3.37  & 6.25  \\
Cu & 2.46  & $\cdots$ & $\cdots$ & 1.94  & 2.46  & $\cdots$ & $\cdots$ & 4.21  \\
Y & 0.63  & $\cdots$ & $\cdots$ & $\cdots$ & 0.97  & $\cdots$ & $\cdots$ & 2.24  \\
Ba & 0.36  & $-$0.41  & $-$1.61  & 0.02  & 0.00  & 0.85  & 0.00  & 2.13  \\
La & $-$0.40  & $\cdots$ & $\cdots$ & $-$0.84  & $\cdots$ & $\cdots$ & $\cdots$ & 1.22  \\
Eu & $-$1.13  & $\cdots$ & $\cdots$ & $\cdots$ & $\cdots$ & $\cdots$ & $\cdots$ & 0.51  \\
 &  &  &  &  &  &  &  &  \\
\hline
 &  &  &  &  &  &  &  &  \\
$[$Fe/H$]$ & $-$1.57  & $-$2.17  & $-$2.72  & $-$2.05  & $-$1.33  & $-$1.57  & $-$2.90  &  \\
 &  &  &  &  &  &  &  &  \\
$[$C/Fe$]$ & $-$0.01  & 0.32  & 0.60  & $-$0.49:  & $-$0.06  & $-$0.05  & $\cdots$ &  \\
$[$O/Fe$]$ & 0.60  & 0.75  & 0.80  & 0.64  & 0.20  & 0.56  & 0.99  &  \\
$[$Na/Fe$]$ & $\cdots$ & $\cdots$ & $\cdots$ & $\cdots$ & $\cdots$ & $\cdots$ & 0.35  &  \\
$[$Mg/Fe$]$ & 0.38  & 0.23  & 0.57  & 0.16  & 0.20  & 0.31  & 0.45  &  \\
$[$Al/Fe$]$ & $-$0.23  & $\cdots$ & $\cdots$ & $\cdots$ & $\cdots$ & $\cdots$ & $\cdots$ &  \\
$[$Si/Fe$]$ & 0.25  & 0.58  & 0.39  & 0.26  & 0.17  & 0.41  & $\cdots$ &  \\
$[$S/Fe$]$ & 0.28  & 0.83  & 1.11:  & 0.52  & 0.19  & 0.34  & $\cdots$ &  \\
$[$K/Fe$]$ & $\cdots$ & $\cdots$ & $\cdots$ & $\cdots$ & $\cdots$ & $\cdots$ & 0.36  &  \\
$[$Ca/Fe$]$ & 0.29  & 0.28  & 0.09  & $-$0.03  & 0.17  & 0.28  & 0.25  &  \\
$[$Sc/Fe$]$ & 0.19  & $\cdots$ & $-$0.16  & 0.16  & 0.17  & 0.08  & $-$0.30  &  \\
$[$Ti/Fe$]$ & 0.09  & 0.55  & 0.10  & 0.00  & 0.17  & 0.05  & 0.16  &  \\
$[$Cr/Fe$]$ & $-$0.13  & $\cdots$ & $-$0.05  & $-$0.30  & $-$0.09  & 0.23  & $-$0.33  &  \\
$[$Mn/Fe$]$ & $-$0.45  & $\cdots$ & $\cdots$ & $\cdots$ & 0.00  & $\cdots$ & $\cdots$ &  \\
$[$Co/Fe$]$ & $-$0.02  & $\cdots$ & $\cdots$ & $-$0.18  & 0.08  & $\cdots$ & $\cdots$ &  \\
$[$Ni/Fe$]$ & $-$0.03  & $-$0.15  & $-$0.11  & $-$0.15  & $-$0.03  & $-$0.06  & 0.02  &  \\
$[$Cu/Fe$]$ & $-$0.18  & $\cdots$ & $\cdots$ & $-$0.22  & $-$0.42  & $\cdots$ & $\cdots$ &  \\
$[$Y/Fe$]$ & $-$0.04  & $\cdots$ & $\cdots$ & $\cdots$ & 0.06  & $\cdots$ & $\cdots$ &  \\
$[$Ba/Fe$]$ & $-$0.20  & $-$0.37  & $-$1.02  & $-$0.06  & $\cdots$ & 0.29  & $\cdots$ &  \\
$[$La/Fe$]$ & $-$0.05  & $\cdots$ & $\cdots$ & $-$0.01  & $\cdots$ & $\cdots$ & $\cdots$ &  \\
$[$Eu/Fe$]$ & $-$0.07  & $\cdots$ & $\cdots$ & $\cdots$ & $\cdots$ & $\cdots$ & $\cdots$ &  \\
 &  &  &  &  &  &  &  &  \\
\hline
\end{tabular}
\end{center}
N{\scshape ote} ---
The solar abundances, with which stellar abundances are compared,
were adopted from Anders \& Grevesse (1989), except for that of Fe 
(7.51) which was taken from Holweger, Kock, \& Bard (1995). Values with large
uncertainties are indicated with colons (:).
\end{table}

\clearpage
\begin{table}[h]
\caption{ O{\scshape xygen} A{\scshape bundance} R{\scshape esults} 
D{\scshape erived} F{\scshape rom} V{\scshape arious} L{\scshape ines}}
\begin{center}
\scriptsize
\begin{tabular}{ccrrrrrc}\hline\hline
line & $\xi$ (${\rm km}\; {\rm s}^{-1}$)& $W_{\lambda}$ (m$\rm\AA$) & $\Delta\log\epsilon$ & $\log\epsilon$ & $\Delta_{T}$ & $\Delta_{g}$ & $<\log\epsilon>_{7773}$ \\
\hline
\multicolumn{2}{c}{HD 44007} & & & & & &  \\
6363 & 1.4 & 4.0 & 0.00  & 7.67  & 0.05 & 0.13 &  \\
8446 & 1.4 & (18.5) & $-$0.10  & 7.98  & $-$0.17 & 0.11 &  \\
7772 & 1.4 & 20.3 & $-$0.14  & 7.88  & $-$0.18 & 0.11 &  \\
7774 & 1.4 & 19.0 & $-$0.13  & 7.99  & $-$0.18 & 0.11 &  \\
7775 & 1.4 & 13.8 & $-$0.12  & 8.01  & $-$0.16 & 0.11 &  \\
 &  &  &  &  &  &  & 7.96  \\
\multicolumn{2}{c}{HD 84937} & & & & & &  \\
6363 & 1.1 & $\cdots$ & $\cdots$ & $\cdots$ & $\cdots$ & $\cdots$ &  \\
8446 & 1.1 & (15.6) & $-$0.07  & 7.50  & $-$0.12 & 0.10 &  \\
7772 & 1.1 & 19.2 & $-$0.10  & 7.45  & $-$0.11 & 0.10 &  \\
7774 & 1.1 & 17.1 & $-$0.09  & 7.54  & $-$0.11 & 0.10 &  \\
7775 & 1.1 & 11.5 & $-$0.08  & 7.55  & $-$0.11 & 0.10 &  \\
 &  &  &  &  &  &  & 7.51  \\
\multicolumn{2}{c}{HD 88609} & & & & & &  \\
6363 & 1.9 & 2.8 & 0.00  & 6.80  & 0.10 & 0.09 &  \\
8446 & 1.9 & $\cdots$ & $\cdots$ & $\cdots$ & $\cdots$ & $\cdots$ &  \\
7772 & 1.9 & 4.6 & $-$0.10  & 7.13  & $-$0.19 & 0.12 &  \\
7774 & 1.9 & 2.0 & $-$0.09  & 6.89  & $-$0.19 & 0.12 &  \\
7775 & 1.9 & $\cdots$ & $\cdots$ & $\cdots$ & $\cdots$ & $\cdots$ &  \\
 &  &  &  &  &  &  & 7.01  \\
\multicolumn{2}{c}{HD 165195} & & & & & &  \\
6363 & 1.8 & 10.9 & 0.00  & 7.47  & 0.05 & 0.13 &  \\
8446 & 1.8 & (3.7) & $-$0.07  & 7.38  & $-$0.22 & 0.11 &  \\
7772 & 1.8 & 7.3 & $-$0.09  & 7.44  & $-$0.20 & 0.12 &  \\
7774 & 1.8 & 5.0 & $-$0.09  & 7.40  & $-$0.20 & 0.13 &  \\
7775 & 1.8 & 6.3 & $-$0.10  & 7.73  & $-$0.20 & 0.12 &  \\
 &  &  &  &  &  &  & 7.52  \\
\multicolumn{2}{c}{HD 175305} & & & & & &  \\
6363 & 1.5 & 2.8 & 0.00  & 7.81  & 0.07 & 0.12 &  \\
8446 & 1.5 & (16.2) & $-$0.09  & 7.82  & $-$0.17 & 0.10 &  \\
7772 & 1.5 & 21.1 & $-$0.13  & 7.81  & $-$0.16 & 0.10 &  \\
7774 & 1.5 & 17.7 & $-$0.12  & 7.86  & $-$0.16 & 0.10 &  \\
7775 & 1.5 & 9.6 & $-$0.11  & 7.73  & $-$0.15 & 0.11 &  \\
 &  &  &  &  &  &  & 7.80  \\
\multicolumn{2}{c}{HD 184266} & & & & & &  \\
6363 & 2.2 & 1.3: & 0.00  & 7.47:  & 0.10 & 0.09 &  \\
8446 & 2.2 & (109.0) & $-$0.26  & 8.06  & $-$0.14 & 0.12 &  \\
7772 & 2.2 & 106.8 & $-$0.46  & 8.17  & $-$0.13 & 0.10 &  \\
7774 & 2.2 & 90.1 & $-$0.36  & 8.16  & $-$0.13 & 0.10 &  \\
7775 & 2.2 & 62.5 & $-$0.30  & 8.06  & $-$0.13 & 0.11 &  \\
 &  &  &  &  & & &  \\
6363 & 5.1 & 1.3: & 0.00  & 7.47:  & 0.10 & 0.09 &  \\
8446 & 5.1 & (109.0) & $-$0.23  & 7.98  & $-$0.13 & 0.11 &  \\
7772 & 5.1 & 106.8 & $-$0.29  & 7.92  & $-$0.13 & 0.11 &  \\
7774 & 5.1 & 90.1 & $-$0.27  & 7.94  & $-$0.12 & 0.11 &  \\
7775 & 5.1 & 62.5 & $-$0.22  & 7.91  & $-$0.12 & 0.11 &  \\
 &  &  &  &  &  &  & 7.92  \\
\multicolumn{2}{c}{CS 22892-052} & & & & & &  \\
6300 & 1.9 & 3.4: & 0.00  & 6.65:  & 0.11 & 0.09 &  \\
6363 & 1.9 & 4.3: & 0.00  & 7.26:  & 0.11 & 0.08 &  \\
7772 & 1.9 & 5.4 & $-$0.10  & 7.02  & $-$0.18 & 0.12 &  \\
 &  &  &  &  &  &  & 7.02  \\
\hline
\end{tabular}
\end{center}
\scriptsize
N{\scshape ote} ---
The $W_{\lambda}$ values for O~{\scshape i} 8446 (blended with Fe~{\scshape i} 
lines) presented here (with parentheses) were inversely computed from the 
abundances derived from profile-fitting, so that they may comprise the 
contribution of oxygen components only.
$\Delta\log\epsilon$ (4th clumn) is the NLTE abundance correction
($\equiv$ $\log\epsilon_{\rm NLTE}-\log\epsilon_{\rm LTE}$),
while $\log\epsilon$ (5th clumn) is the NLTE abundance
($\log\epsilon_{\rm NLTE}$). $\Delta_{T}$ and $\Delta_{g}$ 
(6th and 7th column) are the abundance variations in response
to changes of $\Delta T$ = +150 K and $\log g$ = +0.3, respectively.
$<\log\epsilon>_{7773}$ is the mean NLTE abundance of the 7771--5 triplet
averaged over each of the three components, which is the final oxygen 
abundance adopted in this paper. Regarding HD 184266, two set of results
corresponding to two $\xi$ values [ 2.2 ${\rm km}\; {\rm s}^{-1}$ 
(from Fe~{\sc i} lines) and 5.1 ${\rm km}\; {\rm s}^{-1}$ (from 
O~{\scshape i} triplet) ] are presented, though the latter set was used 
for calculating $<\log\epsilon>_{7773}$.
Values with large uncertainties are indicated with colons (:).
\end{table}

\clearpage



\figcaption[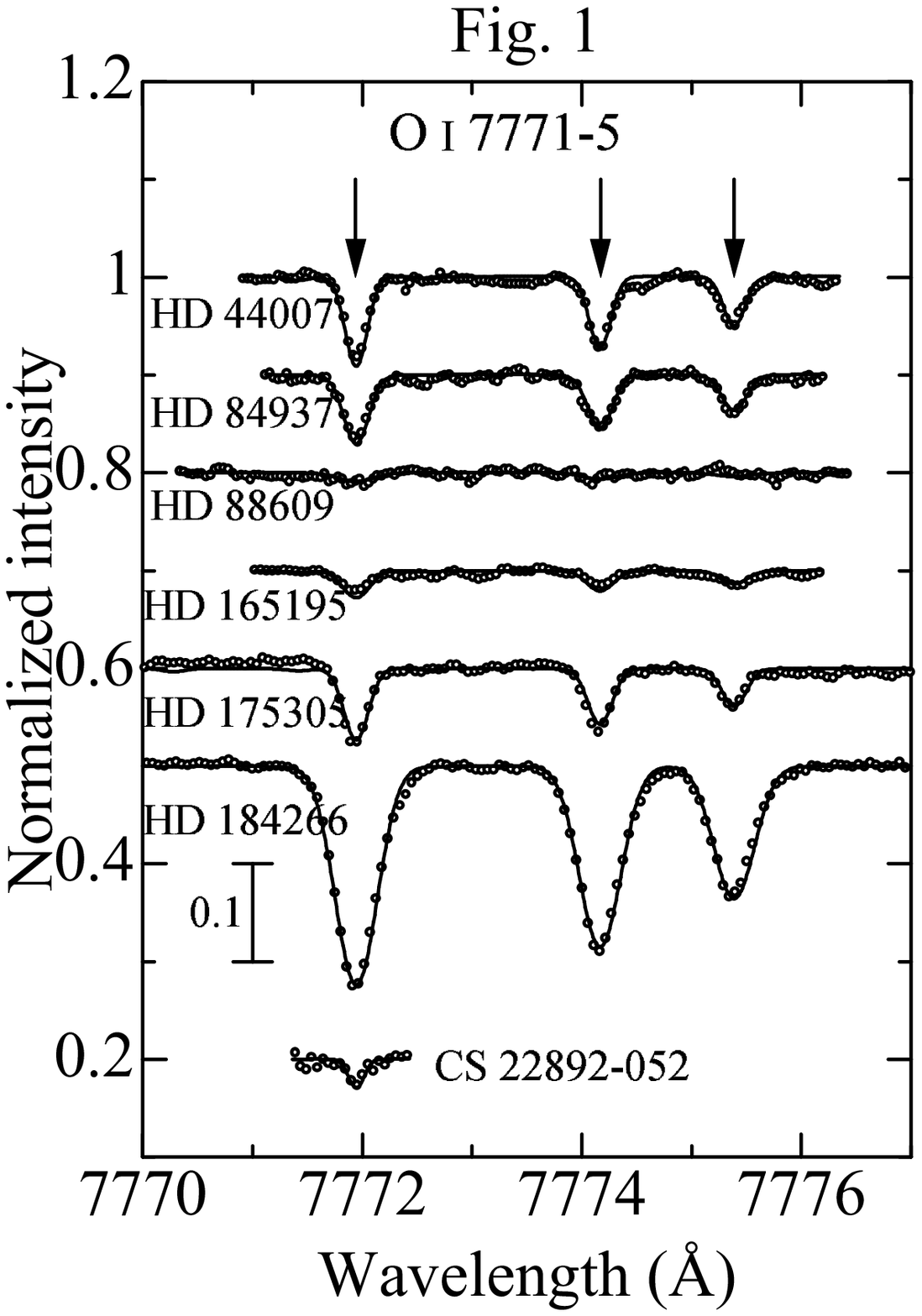]{Spectrum portion comprising the O~{\scshape i} 7771--5
triplet, where observed data are shown by circles. Solid lines
indicate the theoretical spectra which were computed using the
$<\log\epsilon_{7773}>$ [mean abundance over each component of
7771--5 triplet; i.e., the finally adopted oxygen abundance in this paper 
(cf. Table 3)]. The positions of the component lines are indicated by arrows.
An appropriate vertical offset is applied to each spectrum
relative to the adjacent one.}

\figcaption[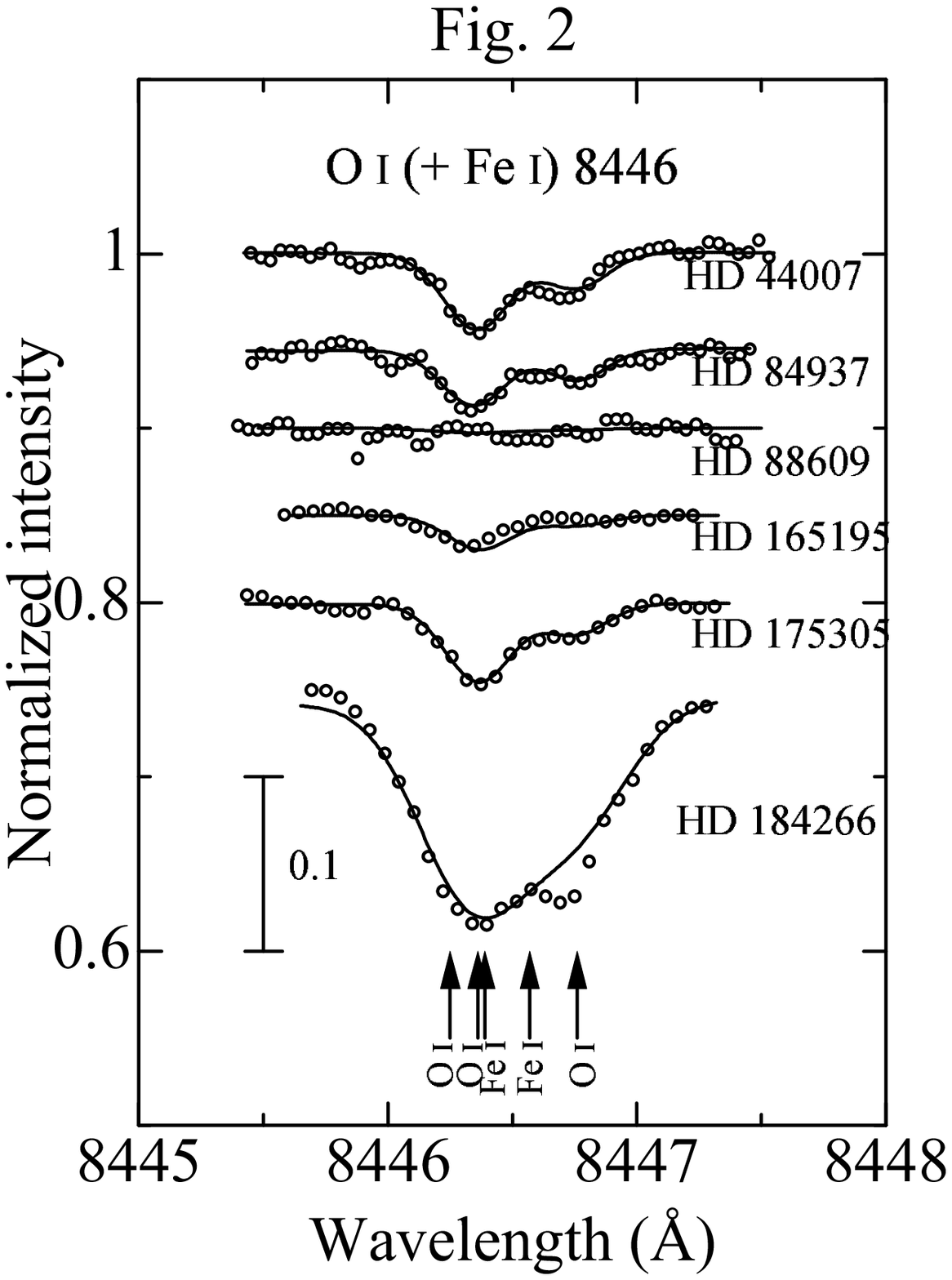]{Spectrum portion comprising the O~{\scshape i} 
(+ Fe~{\scshape i}) 8446 feature. Observed data are shown by circles. 
Solid lines are the theoretical spectra corresponding to 
$<\log\epsilon_{7773}>$ as in Figure 1.}

\figcaption[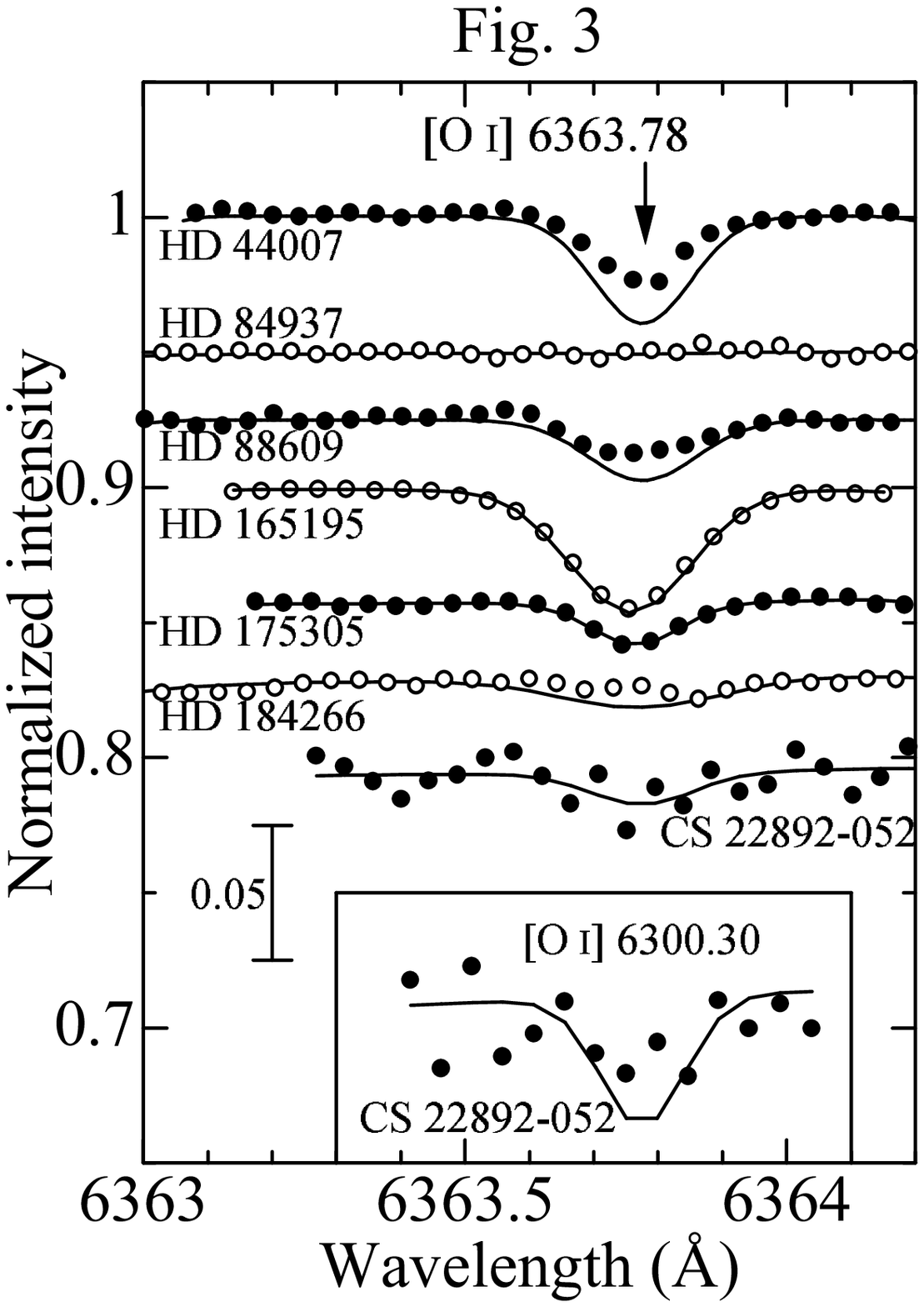]{Spectrum portion at the forbidden line 
[O~{\scshape i}] 6363, where observed data are shown by circles.
(For CS 22892-052, the portion of [O~{\scshape i}] 6300 is additionally shown 
in the inset.)
The theoretical spectra corresponding to the adopted oxygen
abundance ($<\log\epsilon_{7773}>$) are shown by solid lines as in Figure 1.}

\figcaption[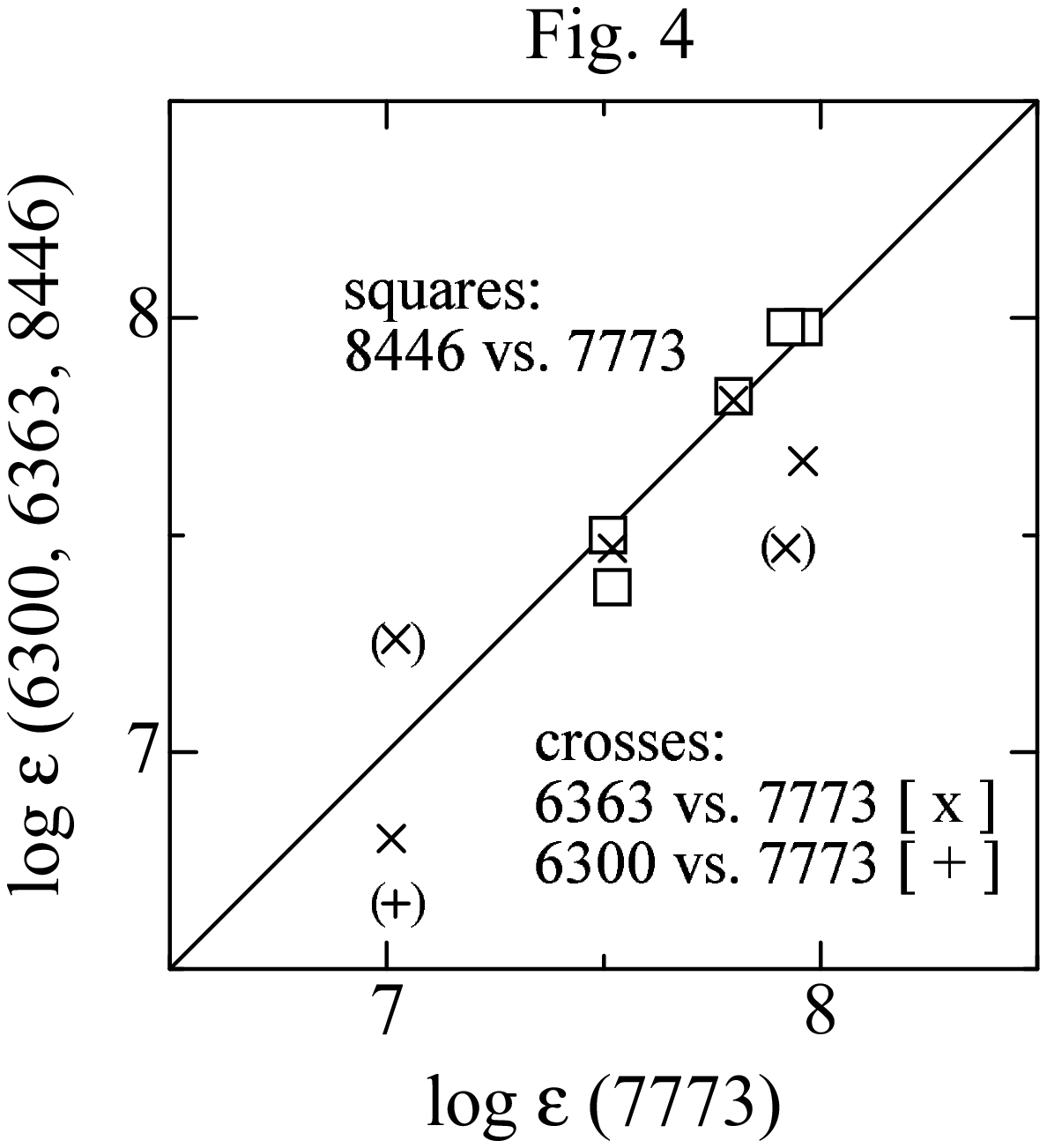]{Comparison of the adopted oxygen abundance 
derived from O~{\scshape i} 7771--5 ($<\log\epsilon_{7773}>$) with that of
O~{\scshape i} 8446 or [O~{\scshape i}] 6363 (6300). Parenthesized symbols indicate the data
with large uncertainties.}

\figcaption[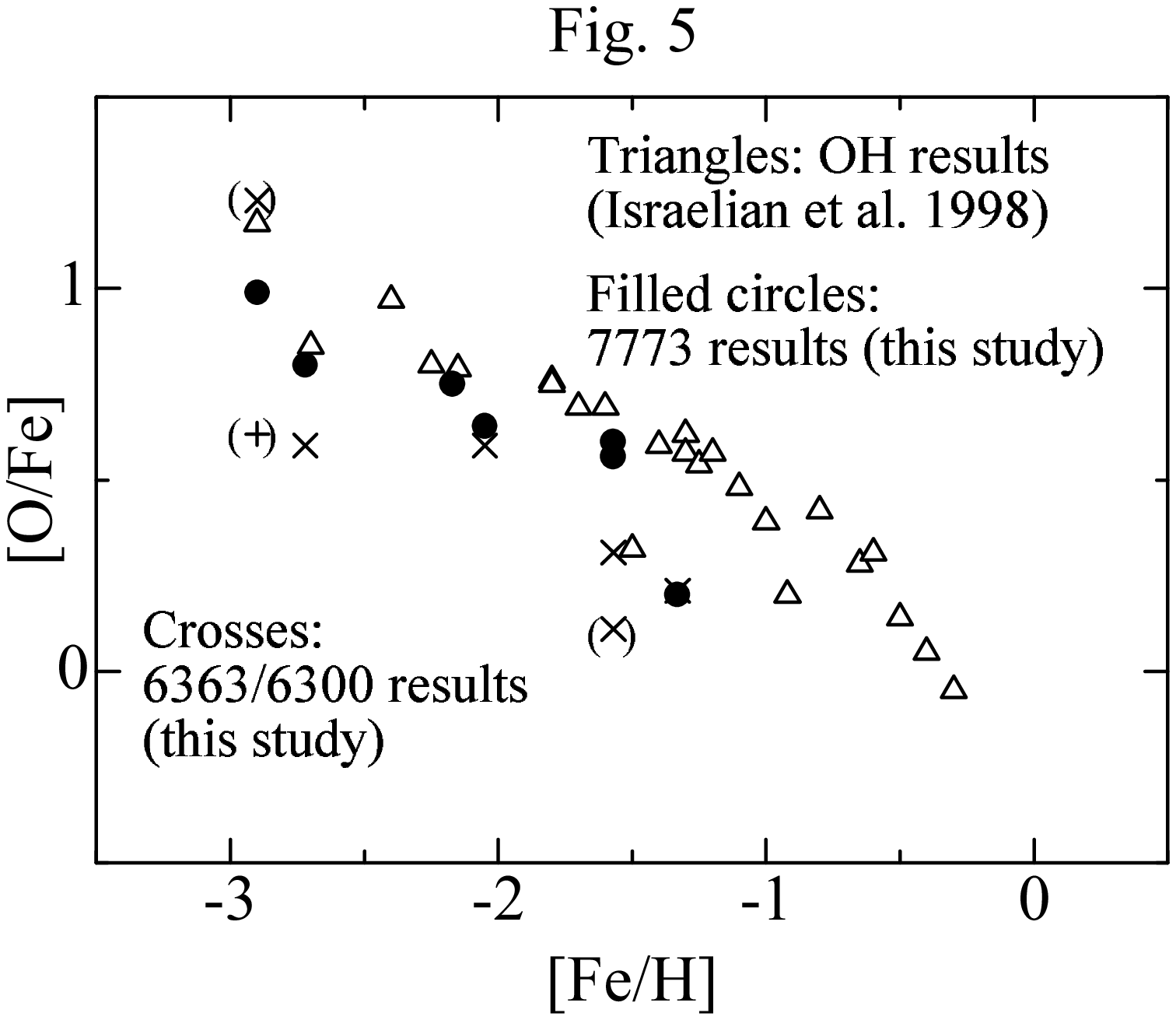]{[O/Fe] vs. [Fe/H] correlation resulting from the
present analysis (the circles represent our O~{\scshape i} 7771-5 results 
finally adopted, while crosses represent the [O~{\scshape i}] results), 
overplotted on the relation derived by Israelian et al. (1998) based on the 
OH lines in UV (triangles). Parenthesized symbols are uncertain values.}

\clearpage
\begin{figure}[h]
\includegraphics[]{figure1.ps}
\end{figure}
\begin{figure}[h]
\includegraphics[]{figure2.ps}
\end{figure}
\begin{figure}[h]
\includegraphics[]{figure3.ps}
\end{figure}
\begin{figure}[h]
\includegraphics[]{figure4.ps}
\end{figure}
\begin{figure}[h]
\includegraphics[]{figure5.ps}
\end{figure}


\end{document}